# HIGH ENERGY SCATTERING OF POLARIZED NUCLEONS WITH THE PHENOMENOLOGICAL SPIN – ORBIT POTENTIAL

Nguyen Nhu Xuan[*]

*Department of Physics, Le Qui Don University, Hanoi, Vietnam*

**Abstract**. The scattering amplitude of polarized nucleons has been found within the framework of the Klein – Gordon with the phenomenological spin - orbit potential. It has the Glauber type representation. The differential cross sections of polarized nucleon are considered and discussed. The Yukawa potential is applied for this problem to determine the polarization of high energy scattering nucleons.

## 1. Introduction

In two papers of H. S. Köhler, Cern, Geneva [8,9], he pointed out several measurements, which have been made of differential cross-sections and polarizations of protons inelastically scattered by nuclei. Such experiments have been made at 220 MeV in Rochester, at 155 and 173 MeV in Uppsala and at 135 and 95 MeV in Harwell. A striking feature of these results is the similarity between the angular dependence of the polarization of particles scattered inelastically by exciting a low-lying level and the elastic polarization. Even more striking is the agreement between polarizations of protons scattered inelastically by nuclei of different masses.

In the paper [10], the basic element for an evaluation of the complex spin-orbit part of the optical potential is the calculation of the complex effective internucleon spin-orbit interaction was considered. Pervious investigators concerned with the real part of the spin-orbit potential, have taken this effective spin-orbit force as shortratige in comparison with the effective forces which give rise to the central part of the optical potential. The reason behind this assumption is the short-range behaviour ofthe spin-orbit component entering in the free internucleon interaction, for example the Hamada-Johnston force. The difference between the effective and the free spin-orbit internucleon force appears mainly near and inside the core region and is due to the presence of the tensor and higher order components in the realistic internucleon force and the Pauli principle. The extrapolation of these results to the nuclear scattering case which we consider does not seem to be straightforward. Similar effects can also be expected to influence the spin-orbit effective interaction.

___________
[*] Corresponding author. Tel.098.3328776,
 email: xuannn@mta.edu.vn

The calculation of the real part of the spin-orbit part of the optical potential for spin saturated nuclei has usually followed the procedure given by Blin-Stoyle. He found that exchange effects account for about half the strength of the spin-orbit component of the potential and therefore cannot be neglected. Again, as remarked earlier, the importance of exchange contributions cannot be readily extrapolated from the bound state case to the nuclear scattering situation since we expect that these effects should decrease as the incident nucleon energy increases [10].

Moreover, the polarizations which result from the elastic scattering of protons by spin-zero nuclei are discussed by means of the multiple-diffraction approximation. They are found to depend largely on the elementary proton-nucleon central-force and spin-orbit scattering amplitudes and, at least for heavy nuclei, to depend very little upon the remaining proton-nucleon amplitudes which are proportional to the spins of the target nucleons. In particular, the values of the polarization measured at small angles can be used to determine the imaginary part of the forward proton nucleon spin-orbit scattering amplitude. The real part of that amplitude, on the other hand, is much less sensitively determined by the polarization distributions. Its evaluation will probably require more detailed measurements on the proton spin distribution. The effects of the Coulomb field on the polarization distributions are significant at all scattering angles. Included among these electromagnetic effects is the relativistic interaction of the proton magnetic moment with the Coulomb field. [11]

Scattering of particles included spin s = ½ considered in some papers [5] – [7], but methods were not complete, or could not be applied for various potentials. In the paper of Kuleshov et al [14] used two – components method to study scattering of particle with spin ½ , but the affection of spin to scattering amplitude was not clearly.

In the recent our paper [23], we have used the Dirac equation in an external field to investigate the Glauber representation for scattering of Dirac particles (spinor particles) in the smooth potential after using the Foldy – Wouthuysen transformation.

This paper aims to generalize the eikonal representation for the scattering amplitude of spinor particles at high energy. We used Klein – Gordon equation in non-relativistic approximation in the form "two-component formalism" to study the scattering of nucleons with spin. But Klein – Gordon is the second order differential respect to time also it can be transformed into two coupled first order differential equations after using "two components formalism. We gain several advantages from of this reduction. First, the equation is now first order equation in the time, so that the time dependence of the two-component wave function is uniquely determined by its initial value, in agreement with the rules of quantum mechanics. Second, the interaction of nucleons with spin ½ will be tied to the 2 x 2 Pauli matrices so the use of the two - component formalism will comment accordingly with spinor matrix structure as in quantum mechanics phenomenology.

Nonrelativistic approximation is used here to separate the classical contribution and the contribution of the spin into the scattering amplitude, from which we can compare with the previous results [23]. Also due to the inclusion of virtual potential phenomenology related to the absorption of virtual nucleon, we consider the more is the polarization of the nucleons in the scattering process.

The paper is organized as follows. In the second section, we obtain the Klein – Gordon equation in an external field in the non-relativistic approximation by using "two-component formalism". In Section 3, we get the scattering amplitude of high energy nucleons with phenomenological spin - orbit potential. Section 4 is devoted to compute the analytical expressions of the differential cross section and polarization of nucleons in the Yukawa potential. The results and possible generalizations of this approach are also discussed in section 5.

**2. Klein – Gordon non-relativistic equation in the external field**

The Klein – Gordon equation in the external field with $A_\mu(x)$ - 4 components potential (q – electric factor included, m is mass of the particle) has form

$$\left[-\left(i\partial_\mu - A_\mu\right)\left(i\partial^\mu - A^\mu\right) + m^2\right]\psi(x) = 0 \tag{2.1}$$

We will now discuss how above equation can be cast into a "two-component" form. This will help to understand the equation and to study its nonrelativistic limit.

Above equation will be cast into two equations in the Schrodinger equation form

$$i\frac{\partial \psi}{\partial t} = H\psi \tag{2.2}$$

where $\phi$ is a vector in a complex two – dimensional space and H is a 2x2 matrix [19].

The transformation to two – component form can be carried out by introducing new two wave functions $\phi_+(r,t), \phi_-(r,t)$, which are more symmetric linear combination of $\psi(r,t)$ and $\frac{\partial \psi(r,t)}{\partial t}$.

$$\phi_+(r,t) = \frac{1}{\sqrt{2m}}\left(i\frac{\partial}{\partial t} - V^0 + m\right)\psi(r,t)$$

$$\phi_-(r,t) = \frac{1}{\sqrt{2m}}\left(-i\frac{\partial}{\partial t} + V^0 + m\right)\psi(r,t) \tag{2.3}$$

The choice (2.3) is not unique. It was chosen because it is simple and gives equations with some features suggestive of the Dirac equation.

With the choice (2.3), now we go into defining a new wave function. It will be organized into a two – component column vector

$$\psi(r,t) = \begin{pmatrix} \phi_+(r,t) \\ \phi_-(r,t) \end{pmatrix}. \tag{2.4}$$

This vector satifies the first order diffential equation (2.2) with Hamintonian is []

$$H = \begin{bmatrix} m + V^0 + \dfrac{(\boldsymbol{p}-\vec{V})^2}{2m} & \dfrac{(\boldsymbol{p}-\vec{V})^2}{2m} \\ -\dfrac{(\boldsymbol{p}-\vec{V})^2}{2m} & -m + V^0 - \dfrac{(\boldsymbol{p}-\vec{V})^2}{2m} \end{bmatrix}$$

$$= \left[ m + \dfrac{(\boldsymbol{p}-\vec{V})^2}{2m} \right] \sigma_3 + \dfrac{(\boldsymbol{p}-\vec{V})^2}{2m} i\sigma_2 + V^0 \tag{2.5}$$

where $\boldsymbol{p} = -i\nabla$ is the energy momentum opeartor. The generalized "potential" interaction consists of a vertor part $\vec{V}$ and a scalar part $V^0$ and $\sigma_i$ are the Pauli matrices.

In our present study, we obtain explicit relations for the scattering of a non-relativistic particle, while we plan to consider the corresponding relativistic problem in the future. Therefore, we assume that the solution of eq.(2.2) has form

$$\psi(r,t) = \begin{pmatrix} \phi_+(r,t) \\ \phi_-(r,t) \end{pmatrix} = \begin{pmatrix} \phi_1(r) \\ \phi_2(r) \end{pmatrix} e^{-iEt} = \begin{pmatrix} \phi_1(r) \\ \phi_2(r) \end{pmatrix} e^{-i(m+T)t} \tag{2.6}$$

where T is the kinetic energy of the particle.

Substituting eq.(2.6) into eq.(2.4) and using Hamiltonian in eq. (2.5), we have

$$(m+T) \begin{pmatrix} \phi_1(r) \\ \phi_2(r) \end{pmatrix} = \begin{bmatrix} m + V^0 + \dfrac{(\boldsymbol{p}-\vec{V})^2}{2m} & \dfrac{(\boldsymbol{p}-\vec{V})^2}{2m} \\ -\dfrac{(\boldsymbol{p}-\vec{V})^2}{2m} & -m + V^0 - \dfrac{(\boldsymbol{p}-\vec{V})^2}{2m} \end{bmatrix} \begin{pmatrix} \phi_1(r) \\ \phi_2(r) \end{pmatrix} \tag{2.7}$$

Equation (2.7) can be seperated into coupled equations as follow

$$\begin{cases} T\phi_1(r) = \left[ V^0 + \dfrac{(\boldsymbol{p}-\vec{V})^2}{2m} \right] \phi_1(r) + \dfrac{(\boldsymbol{p}-\vec{V})^2}{2m} \phi_2(r) \\ (2m+T)\phi_2(r) = \left[ -\dfrac{(\boldsymbol{p}-\vec{V})^2}{2m} \right] \phi_1(r) - \left[ \dfrac{(\boldsymbol{p}-\vec{V})^2}{2m} - V^0 \right] \phi_2(r) \end{cases} \tag{2.8}$$

As $m \to \infty$, the dimensionless quantities $\frac{|p|}{m}, \frac{|\vec{V}|}{m}$ and $\frac{|T|}{m}$ are all $\ll 1$ therefore $\phi_2(r) \ll \phi_1(r)$.

Expanding the second equation (2.7) in inverse powers of $m$ and discarding terms of order $m^{-3}$ or higher gives

$$\phi_2(r) \simeq -\frac{(p-\vec{V})^2}{4m^2}\phi_1(r) + O\left(\frac{1}{m^3}\right) \tag{2.9}$$

Substituting this result into the first equation of eq.(2.8) gives an equation for $\phi_1$ accurate to order $m^{-3}$

$$T\phi_1(r) = \left[\frac{1}{2m}(p-\vec{V})^2 + V^0 - \frac{1}{8m^3}(p-\vec{V})^4\right]\phi_1(r) \tag{2.10}$$

In the specific case, when the external field is scalar, $\vec{V} = 0$, eq. (2.10) becomes:

$$T\phi_1(r) = \left[-\frac{\nabla^2}{2m} + V^0 - \frac{\nabla^4}{8m^3}\right]\phi_1(r) \tag{2.11}$$

The right hand side of eq.(2.11) has two terms, the first term $\left[-\frac{\nabla^2}{2m} + V^0\right]$ is the non-relativistic limit part, and the second term $\left[-\frac{\nabla^4}{8m^3}\right]$ is the relativistic correction to the energy up to order $\frac{1}{m^3}$.

Now, we drop out the second term and retain the first term of RHS of eq.(2.1), and note that $T = \frac{p^2}{2m}$, we obtain

$$\left(p^2 + \nabla^2\right)\phi_1(r) = 2mV^0\phi_1(r) \quad . \tag{2.12}$$

Set $V(r) = 2mV^0$, then

$$\left(p^2 + \nabla^2\right)\phi_1(r) = V(r)\phi_1(r) \tag{2.13}$$

Thus, from two - component formalism, we also obtain the Klein – Gordon (KG) non-relativistic equation with *V(r)* potential..

Since the KG equation describes the behavior of a spin zero particle therefore it would be a good approximation equation to describe pionic atoms (atomic states with π⁻ substituted for an electron). Knowing that the wave function of the Dirac particle also satisfies the KG equation, we can apply the eq.(2.12) for the Dirac particle noting that $\phi_1$ is a spinor having two - components and V is an 2x2 matrix operator having spin interaction.

**3. High energy scattering amplitude of nucleons in smooth interaction potential**

In nuclear physics, the elastic scattering of nucleons in nuclear can be described by using spin – orbit potential [18]

$$V(r,\vec{\sigma}) = -(1+i\xi)U(r) + \frac{a}{r}\frac{dU(r)}{dr}\vec{\sigma}.\vec{L} \qquad (3.1)$$

where $\vec{L} = -i[\vec{r} \times \nabla]$ is angular momentum operator, a – constant has square length dimension and the imaginary part $i\xi U(r)$ of this potential is considered due to the absorption nucleon by nuclear.

Set
$$V_0 = -(1+i\xi)U(r); V_1 = \frac{a}{r}\frac{dU(r)}{dr} \qquad (3.2)$$

then
$$V(r) = V_0(r) + V_1(r)(\vec{\sigma}.\vec{L}), \qquad (3.3)$$

For smooth potential, the quasi-classical condition of scattering is satisfied [17, 30]

$$\left|\frac{\dot{V}}{V_p}\right| \ll 1, \left|\frac{\dot{V}}{p^2}\right| \ll 1 \qquad (3.4)$$

The solutions of equations (2.13) with the boundary conditions (3.4) can be written in the form

$$\phi_1(r) = \varphi(r).e^{ipz} \qquad (3.5)$$

where $\varphi(r)$ is a Dirac particle having two components, with the boundary conditions $\varphi(r)|_{z \to -\infty} = \varphi_0 \equiv \varphi_{1/2,m_s}$. Here, $\varphi_{1/2,m_s}$ are spin function [29]:

$$\varphi_{1/2,1/2} = \begin{pmatrix} 1 \\ 0 \end{pmatrix} \text{ and } \varphi_{1/2,-1/2} = \begin{pmatrix} 0 \\ 1 \end{pmatrix} \qquad (3.6)$$

Substituting eq.(3.5) into eq.(2.13), we have

$$2ipe^{ipz}\frac{\partial \varphi(r)}{\partial z} + e^{ipz}\nabla^2 \varphi(r) = V(r)e^{ipz}\varphi(r) \qquad (3.7)$$

where $\vec{r} = (\vec{b}, z)$, with the condition (3.4), the Dirac particle φ(r) are slowly varying functions and the z-axis is chosen to be coincident with the direction of incident momentum $\vec{p}$.

Thus, φ(r) approximately satisfy the equation

$$2ipe^{ipz}\frac{\partial \varphi(r)}{\partial z} = \left[V_0(r) - iV_1(r)\vec{\sigma}[\vec{r} \times \nabla]\right]e^{ipz}\varphi(r) \qquad (3.8)$$

To perform some calculations and retain the p-terms and use the boundary condition, we obtain

$$\ln \varphi = \ln \varphi_0 + \frac{1}{2ip} \int_{-\infty}^{z} \left[ V_0 + V_1 p [\vec{\sigma} \times \vec{r}]_z \right] dz'$$

$$\Leftrightarrow \varphi = \varphi_0 \exp\left\{ \frac{1}{2ip} \int_{-\infty}^{z} \left[ V_0(b, z') + V_1 p [\vec{\sigma} \times \vec{r}]_z \right] dz' \right\}$$

(3.9)

Here, $\vec{r} = (\vec{b}, z)$; $\vec{n} = \frac{\vec{b}}{b} = (\cos\theta, \sin\theta)$, where $\theta$ is the azimuthal angle in the $(x, y)$ - plane.

Equation (3.8) can be rewritten as

$$\varphi = \varphi_0 \exp\left\{ \frac{1}{2ip} \int_{-\infty}^{z} V_0(b, z') dz' + i \frac{b[\vec{n} \times \vec{\sigma}]_z}{2} \int_{-\infty}^{z} V_1(b, z') dz' \right\}$$

(3.10)

Set

$$\chi_0(\vec{b}, z) = \frac{1}{2ip} \int_{-\infty}^{z} V_0(\vec{b}, z') dz'$$

(3.11)

$$\chi_1(\vec{b}, z) = \frac{b}{2} \int_{-\infty}^{z} V_1(\vec{b}, z') dz'$$

(3.12)

Eq. (3.9) rewritten as

$$\varphi(r) = \varphi_0 \cdot \exp\left\{ \chi_0(b, z) + i (\vec{n} \times \vec{\sigma})_z \chi_1(b, z) \right\}$$

(3.13)

Therefore, the solution of eq.(2.13) has the form

$$\phi_1(r) = e^{ipz} \varphi(r) = e^{ipz} \varphi_0 \cdot \exp\left\{ \chi_0(b, z) + i (\vec{n} \times \vec{\sigma})_z \chi_1(b, z) \right\}$$

(3.14)

For the scattering amplitude, one obtains the Glauber representation

$$f(p, \Delta) = -\frac{1}{4\pi} \int dr e^{-ip'r} \varphi_0^*(p') \left( V_0(r) + V_1 p [\sigma \times r]_z \right) \phi_1(r)$$

$$= \frac{p}{2i\pi} \int d^2 b e^{-ib\Delta} \varphi_0^*(p') \left[ e^{\chi_0 + i[n \times \sigma]_z \chi_1} - 1 \right] \varphi_0(p)$$

(3.15)

Note that $[\vec{n} \times \vec{\sigma}] = \sigma_y \cos\theta - \sigma_x \sin\theta$; $\int d^2 \vec{b} ... = \int_0^{2\pi} d\theta \int b db ...$

One can rewrite this formula as

$$f(\theta) = f(p, \Delta) = \varphi_0^*(\vec{p}') \left[ A(\theta) + \sigma_y B(\theta) \right] \varphi_0(\vec{p})$$

(3.16)

$$\Delta = \vec{p}' - \vec{p} = 2p \sin\frac{\theta}{2}; \chi_0 = \chi_0(\vec{b}, \infty), \chi_1 = \chi_1(\vec{b}, \infty)$$

(3.17)

$$A(\theta) = -ip\int_0^\infty bdb J_0(b\Delta)\left[e^{\chi_0}\cos\chi_1 - 1\right] \tag{3.18}$$

$$B(\theta) = -ip\int_0^\infty bdb J_1(b\Delta) e^{\chi_0}\sin\chi_1 \tag{3.19}$$

where $p'$ and $\theta$ are the momentum after scattering and the scattering angle; $J_0(b\Delta)$ and $J_1(b\Delta)$ are the Bessel functions of the zeroth and the first order. The presence of quantities $A(\theta)$ and $B(\theta)$ determined by formulas (3.18) and (3.19) in the high – energy limit shows that there are both spin-flip and non-spin flip parts contributing to the scattering amplitude.

### 4. Differential scattering cross section of polarized nucleon

Square of scattering amplitude (3.16) determines the differential scattering cross section of polarized nucleons. If they are not polarized, the differential scattering cross section is determined by taking the average of two polarized nucleon states $m_s = 1/2, -1/2$ then

$$\frac{d\sigma}{d\Omega} = \frac{1}{2}\sum_{m_s}|f(\theta)|^2 = |A(\theta)|^2 + |B(\theta)|^2 \tag{4.1}$$

The results which relative to scattering of non-polarized nucleon are studied by us in the paper [23] basing on Foldy-Wouthuysen representation apply to Dirac equation.

Now let us examine the scattering of polarized nucleons. Assuming nucleon motions along the y-axis and the spin projection along the z-axis direction. If the nucleons in the plane Oxy is deflected upward positive direction of the z-axis then unit vector that perpendicular to the scattering plane will orientate to the positive direction of the z-axis, so that $\vec{n}.\vec{\sigma} = \sigma_z$. In contrast, the nucleon is skewed to negative z-axis direction, then unit vector perpendicular to the scattering plane will rotate in the negative z-axis direction so $\vec{n}.\vec{\sigma} = -\sigma_z$.

We have the differential scattering cross sections of left and right polarized nucleons respectively

$$\left(\frac{d\sigma}{d\Omega}\right)_{left} = \sum_{m_s}\left|f_{\sigma_z}(\theta)\right|^2 = |A(\theta) + B(\theta)|^2 \tag{4.2}$$

$$\left(\frac{d\sigma}{d\Omega}\right)_{right} = \sum_{m_s}\left|f_{-\sigma_z}(\theta)\right|^2 = |A(\theta) - B(\theta)|^2 \tag{4.3}$$

The polarization of the nucleons has been characterized by polarization vector [27]

$$\vec{P}(\theta) = \vec{n}.\frac{|A^*(\theta).B(\theta)+A(\theta).B^*(\theta)|}{|A(\theta)|^2+|B(\theta)|^2} \tag{4.4}$$

In this section, we use Yukawa potential to compute above differential scattering cross sections and plot graphical of them following the momentum of incident particle and the small scattering angle. Our aim is to compare the influence of the spin and the imaginary part of potential to the polarization of the scattering nucleons.

The Yukawa potential [21] given by:

$$U(r) = \frac{g}{r}e^{-\mu r} = \frac{g}{r}e^{-\frac{r}{R}}, \tag{4.5}$$

here, g is a magnitude scaling constant whose dimension is of energy, $\mu$ is another scaling constant which is related to $R$ - the effective size where the potential is non-zero – as $\mu = \frac{1}{R}$.

Now, we compute the expression $\chi_0(b)$ following eq. (3.11)

$$\chi_0(b,\infty) = \frac{1}{2ip}\int_{-\infty}^{\infty} V_0(\vec{b},z')dz' = -\frac{1+i\xi}{2ip}\int_{-\infty}^{\infty} U(\vec{b},z')dz'$$
$$= -\frac{g(1+i\xi)}{2ip}\int_{-\infty}^{\infty} \frac{e^{-\mu r'}}{r'}dz' = -\frac{g(1+i\xi)}{2ip}\int_{-\infty}^{\infty} \frac{e^{-\mu\sqrt{b^2+z'^2}}}{\sqrt{b^2+z'^2}}dz' \tag{4.6}$$

Using property of the Macdonald function of zeroth order [34]:

$$K_0(\mu b) = \frac{1}{2\pi}\int_0^{\infty} \frac{e^{-\mu r}}{r}dz' = \frac{1}{2\pi}\int_0^{\infty} \frac{e^{-\mu\sqrt{b^2+z'^2}}}{\sqrt{b^2+z'^2}}dz' \tag{4.7}$$

We have

$$\chi_0(b) = -\frac{g(1+i\xi)}{2ip}.(2\pi).K_0(\mu b) = \frac{\pi g(i-\xi)}{p}K_0(\mu b) \tag{4.8}$$

Now, we turn to the calculation of $\chi_1(b)$ following eq. (3.12)

Since:

$$\frac{1}{r}\frac{dU}{dr} = \frac{g}{r}\frac{d}{dr}\left(\frac{e^{-\mu r}}{r}\right) = \frac{-g(1+\mu r)e^{-\mu r}}{r^3} \tag{4.9}$$

We can rewrite $\chi_1(b)$ as

$$\chi_1(b) \equiv \chi_1(b,\infty) = \frac{b}{2}\int_0^\infty V_1(\vec{b},z')dz'$$
$$= \frac{b}{2}.a\int_0^\infty \frac{1}{r}\frac{dU(r)}{dr}dz' = \frac{b}{2}.a\int_0^\infty \frac{-g(1+\mu r)e^{-\mu r}}{r^3}dz' \quad (4.10)$$

with following property

$$\frac{d}{db}(K_0(\mu b)) = \frac{1}{2\pi}\frac{d}{db}\int_0^\infty \frac{e^{-\mu r}}{r}dz' = \frac{1}{2\pi}\int_0^\infty \frac{d}{dr}\left(\frac{e^{-\mu r}}{r}\right).\frac{dr}{db}dz'$$
$$= -\frac{1}{2\pi}\int_0^\infty \left(\frac{e^{-\mu r}+\mu r e^{-\mu r}}{r^2}\right).\frac{b}{r}dz' = -\frac{b}{2\pi}\int_0^\infty \left(\frac{1+\mu r}{r^3}\right)e^{-\mu r}dz' \quad (4.11)$$

one gets:

$$\chi_1(b) = \frac{ga}{2}.2\pi.\frac{d}{db}(K_0(\mu b)) = -\mu ga\pi.K_1(\mu b) \quad (4.12)$$

where $K_1(\mu b)$ is the Macdonald function of first order

$$K_1(\mu b) = -\frac{1}{\mu}\frac{d}{db}(K_0(\mu b)) \quad (4.13)$$

Substitution of eqs. (4.8), (4.13) into eq. (3.18) and (3.19), one gets:

$$A(\theta) = -ip\int_0^\infty bdbJ_0(b\Delta)\left\{\exp\left(\frac{\pi g(i-\xi)}{p}K_0(\mu b)\right).\cos(-\mu ga\pi.K_1(\mu b))-1\right\}$$
$$\simeq -ip\int_0^\infty bdbJ_0(b\Delta).\left(\frac{\pi g(i-\xi)}{p}K_0(\mu b)\right).\cos(-\mu ga\pi.K_1(\mu b)) \quad (4.14)$$
$$\approx \pi g(1+i\xi)\int_0^\infty bdbJ_0(b\Delta)K_0(\mu b) = \frac{\pi g(1+i\xi)}{\Delta^2+\mu^2}$$

$$B(\theta) = -ip\int_0^\infty bdbJ_1(b\Delta)\exp\left(\frac{\pi g(i-\xi)}{p}K_0(\mu b)\right).\sin(-\mu ga\pi.K_1(\mu b))$$
$$\simeq ip\int_0^\infty bdbJ_1(b\Delta)\left[1+\frac{\pi g(i-\xi)}{p}K_0(\mu b)\right]\left[-\mu ga\pi.K_1(\mu b)\right] \quad (4.15)$$
$$\simeq -ip.\mu ga\pi\int_0^\infty bdbJ_1(b\Delta)K_1(\mu b) = -\frac{i\pi agp\Delta}{\mu^2+\Delta^2}$$

The differential scattering cross sections of left and right polarized nucleons follow equations (4.1), (4.2), respectively are

$$\left.\frac{d\sigma}{d\Omega}\right|_{left} = |A(\theta)+B(\theta)|^2 = \left|\frac{\pi g(1+i\xi)}{\Delta^2+\mu^2} - \frac{i\pi agp\Delta}{\mu^2+\Delta^2}\right|^2$$

$$= \left(\frac{\pi g}{\Delta^2+\mu^2}\right)^2 \left[1+(\xi-ap\Delta)^2\right] = \frac{\pi^2 g^2}{\left(4p^2\sin^2\left(\frac{\theta}{2}\right)+\mu^2\right)^2}\left[1+\left(\xi-2ap^2\sin\frac{\theta}{2}\right)^2\right] \quad (4.16)$$

$$\left.\frac{d\sigma}{d\Omega}\right|_{right} = |A(\theta)-B(\theta)|^2 = \left|\frac{\pi g(1+i\xi)}{\Delta^2+\mu^2} + \frac{i\pi agp\Delta}{\mu^2+\Delta^2}\right|^2$$

$$= \left(\frac{\pi g}{\Delta^2+\mu^2}\right)^2 \left[1+(\xi+ap\Delta)^2\right] = \frac{\pi^2 g^2}{\left(4p^2\sin^2\left(\frac{\theta}{2}\right)+\mu^2\right)^2}\left[1+\left(\xi+2ap^2\sin\frac{\theta}{2}\right)^2\right] \quad (4.17)$$

With a dimensionless q defined as $q = \frac{p}{\mu}$ [29], one can rewrite (4.16) and (4.17), respectively, as

$$\left.\frac{d\sigma}{d\Omega}\right|_{left} = \left(\frac{\pi g}{\mu^2}\right)^2 \frac{1+\left(\xi-2a\mu^2 q^2 \sin\frac{\theta}{2}\right)^2}{\left(4q^2\sin^2\left(\frac{\theta}{2}\right)+1\right)^2} \quad (4.18)$$

$$\left.\frac{d\sigma}{d\Omega}\right|_{right} = \left(\frac{\pi g}{\mu^2}\right)^2 \frac{1+\left(\xi+2a\mu^2 q^2 \sin\frac{\theta}{2}\right)^2}{\left(4q^2\sin^2\left(\frac{\theta}{2}\right)+1\right)^2} \quad (4.19)$$

The total differential scattering cross section of nucleon is

$$\frac{d\sigma}{d\Omega} = \frac{1}{2}\left(\left.\frac{d\sigma}{d\Omega}\right|_{left} + \left.\frac{d\sigma}{d\Omega}\right|_{right}\right) = \left(\frac{\pi g}{\mu^2}\right)^2 \frac{1+\xi^2+4a^2\mu^4 q^4 \sin^2\frac{\theta}{2}}{\left(4q^2\sin^2\left(\frac{\theta}{2}\right)+1\right)^2} \quad (4.20)$$

The polarizaton vector of nucleon is

$$\vec{P}(\theta) = \vec{n} \cdot \frac{\frac{\pi g(1-i\xi)}{\Delta^2 + \mu^2} \cdot \left(-\frac{i\pi a g p \Delta}{\mu^2 + \Delta^2}\right) + \frac{\pi g(1+i\xi)}{\Delta^2 + \mu^2} \cdot \left(\frac{i\pi a g p \Delta}{\mu^2 + \Delta^2}\right)}{\frac{d\sigma}{d\Omega}}$$

$$= \vec{n} \left(\frac{\pi g}{\Delta^2 + \mu^2}\right)^2 \cdot a p \Delta \left|-i(1-i\xi) + i(1+i\xi)\right| : \left\{\left(\frac{\pi g}{\Delta^2 + \mu^2}\right)^2 \left[1 + \xi^2 + a^2 p^2 \Delta^2\right]\right\} \quad (4.21)$$

$$= \vec{n} \frac{2 a p \Delta \cdot \xi)}{1 + \xi^2 + a^2 p^2 \Delta^2} = \vec{n} \frac{4\xi a p^2 \sin\theta/2}{1 + \xi^2 + 4 a^2 p^4 \sin^2\theta/2}$$

From eq.(4.21), we derive the maximum value of the polarization vector at scattering angle, that satisfies the condition

$$\sin\theta = \sqrt{\frac{1+\xi^2}{4a^2 p^2}}, \text{ (condition } \sqrt{1+\xi^2} < 2ap\text{)} \quad (4.22)$$

and it depend on ratio of the real and imaginary part of potential.

Evidently, at the high momentum with the imaginary part of potential is small, the nucleon is almost not polarization.

Substituting the scattering angle in eq. (1.22) into eqs.(4.18) – (4.20), we obtain the differential scattering cross sections of left and right polarized nucleons

$$\left.\frac{d\sigma}{d\Omega}\right|_{left} = (\pi g)^2 \frac{1 + \left(\xi - \mu q\sqrt{1+\xi^2}\right)^2}{\left(\frac{1+\xi^2}{a^2} + \mu^2\right)^2} \; ; \; \left.\frac{d\sigma}{d\Omega}\right|_{right} = (\pi g)^2 \frac{1 + \left(\xi + \mu q\sqrt{1+\xi^2}\right)^2}{\left(\frac{1+\xi^2}{a^2} + \mu^2\right)^2} \quad (4.23)$$

$$\frac{d\sigma}{d\Omega} = (\pi g)^2 \frac{(1+\xi^2)(1+\mu^2 q^2)}{\left(\frac{1+\xi^2}{a^2} + \mu^2\right)^2} \quad (4.24)$$

The dependence of the left, right and total differential cross section on q (or, in other words, on the incident momentum) and the scattering angle $\theta$ are graphically plotted in Figures 4.1 and 4.2 (constants are set to unit).

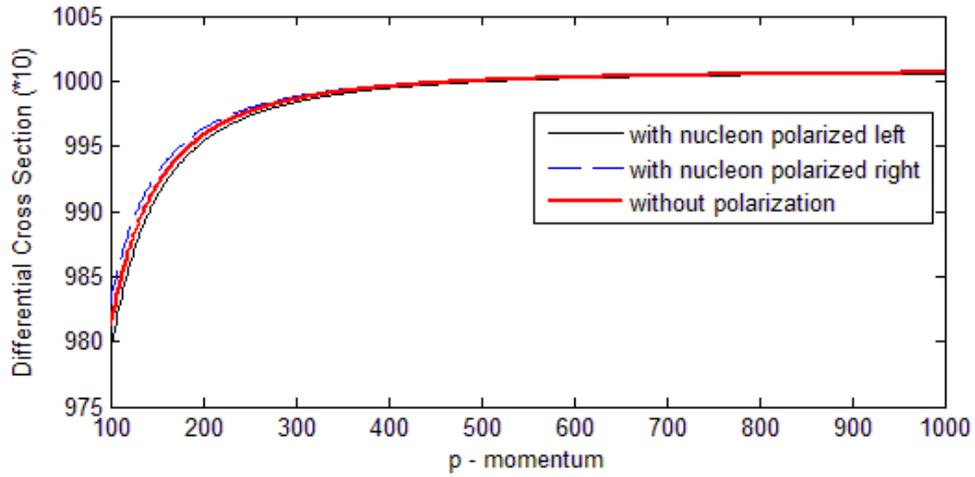

**Fig.4.1.** Dependence of the differential cross section (left, right and total) on the momentum of incident particle (with a specific small value of the scattering angle $\theta = 0.1\,\text{rad}$)

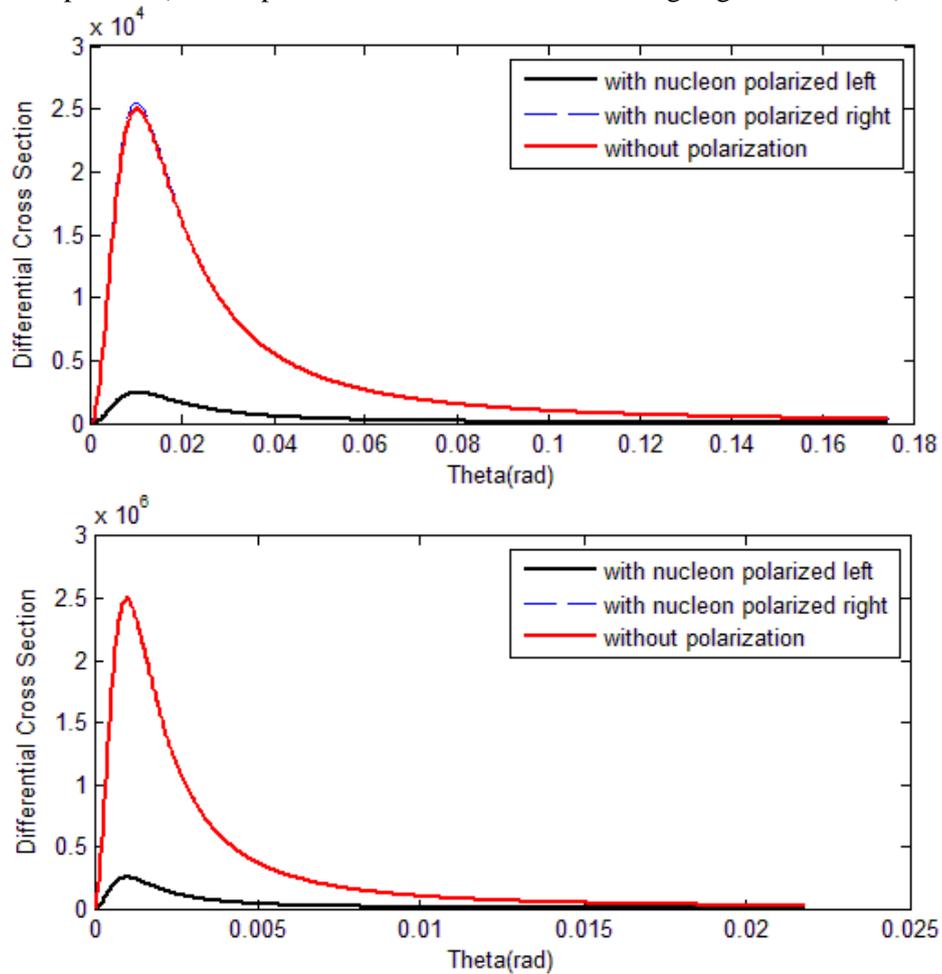

**Fig. 4.2.** Dependence of the differential cross section on the scattering angle for *q = 100 and q = 1000*

In figure 4.2, we show that when the moment incident of particle is high or very high, then the contribution to the scattering process of the left polazired nucleon is small compare with the right polarized nucleon.

## 5. Conclusion

We obtained the nonrelativistic expression for high energy small angles scattering amplitude of nucleons in the form of Glauber type representation.

The expressions for the left, right and total differential cross section for polarization nucleon are obtained by using Klein – Gordon equation with phenomenological spin - orbit potential. The Yukawa potential is used to illustrate these results.

We know that the left polarized nucleon has relatively significant contribution at some finite ranges of incident particle's momentum $p$, however they can be ignored in high energy zone.

if the imaginary part of interaction potential is ignored, the results obtained for the scattering amplitudes of the nucleon feels like the result that we have obtained in recent articles [23].

The problem requires further investigation and that conclusions regarding nuclear structure might then be obtainable.